\documentclass[a4paper]{jpconf}
\usepackage{graphicx}
\usepackage{epsfig}
\usepackage[fleqn,sumlimits,intlimits]{amsmath}
\usepackage{latexsym,amssymb}
\usepackage[english]{babel}
\usepackage{verbatim}
\usepackage{color}
\usepackage{varioref}

\newcommand{\bea}{\begin{eqnarray}}
\newcommand{\eea}{\end{eqnarray}}
\newcommand{\beq}{\begin{equation}}
\newcommand{\eeq}{\end{equation}}

\newcommand{\bal}{\begin{align}}
\newcommand{\eal}{\end{align}}
\newcommand{\bit}{\begin{itemize}}
\newcommand{\eit}{\end{itemize}}

\newcommand{\rar}{\rightarrow}

\newcommand{\abs}[1]{\vert #1\vert}
\newcommand{\avg}[1]{\left\langle #1 \right\rangle}

\newcommand{\dS}{{d_s}}

\begin{document}
\title{Aspects of dynamical dimensional reduction in multigraph ensembles of CDT}

\author{Georgios Giasemidis$^1$, John F. Wheater$^1$ and Stefan Zohren$^{2, 1}$}

\address{$^1$ Rudolf Peierls Centre for Theoretical Physics, 1 Keble Road, Oxford OX1 3NP, UK}
\address{$^2$ PUC-Rio, R. Marqu\^es de S\~ao Vicente 225, Rio de Janeiro, Brazil}

\ead{giasemidis@physics.ox.ac.uk, j.wheater1@physics.ox.ac.uk, zohren@fis.puc-rio.br}

\begin{abstract}
We study the continuum limit of a ``radially reduced'' approximation of Causal Dynamical Triangulations (CDT), so-called multigraph ensembles, and explain why they serve as realistic toy models to study the dimensional reduction observed in numerical simulations of four-dimensional CDT. We present properties of this approximation in two, three and four dimensions comparing them with the numerical simulations and pointing out some common features with 2+1 dimensional Ho\v rava-Lifshitz gravity.
\end{abstract}

\section{Introduction}
The phenomenon of the spectral dimensional reduction has attracted special interest within the quantum gravity community during the last seven years. It was first observed in computer simulations of the CDT (Causal Dynamical Triangulation) approach to quantum gravity \cite{Ambjorn:1998xu} where a scale dependent spectral dimension varying from 4 in the IR to 2 in the UV limit  was observed \cite{Ambjorn:2005db}. The running of the spectral dimension serves as a dynamical mechanism that might regulate the theory at short scales so that it might possesses a non-Gaussian fixed point as suggested by Weinberg \cite{weinberg}. 
Thereafter several authors confirmed a similar result from different approaches to quantum gravity, e.g. asymptotically safe Quantum Einstein Gravity \cite{Litim:2003vp, Lauscher:2005qz}, Ho\v rava-Lifshitz gravity \cite{Horava:2009uw, Horava:2009if}, spin foams \cite{Modesto:2008jz, Caravelli:2009gk, Magliaro:2009if, Modesto:2009qc}, multi-fractional space-times \cite{Calcagni:2010pa, Calcagni:2012rm}. The asymptotic safety scenario and Ho\v rava-Lifshitz gravity share further characteristics with CDT and their relation is an active research field \cite{Reuter:2011ah, Ambjorn:2010hu, Anderson:2011bj}. Besides the contributions from different approaches, within the CDT approach, analytical computation explaining the numerical results seemed out of reach due to the complexity of the problem. To remedy this, the authors recently introduced a radial approximation of CDT, a so-called multigraph ensemble, and showed that it serves as realistic toy model to study dynamical reduction of the spectral dimension in CDT \cite{Giasemidis:2012rf, Giasemidis:2012qk}. In the following we review the results of \cite{Giasemidis:2012rf, Giasemidis:2012qk} and comment on the relation between the radial approximation of CDT and three-dimensional Ho\v rava-Lifshitz gravity.

\section{The multigraph approximation: Motivation and validity}
\label{multigraph_approx}

\begin{figure} 
  \includegraphics[scale=0.3]{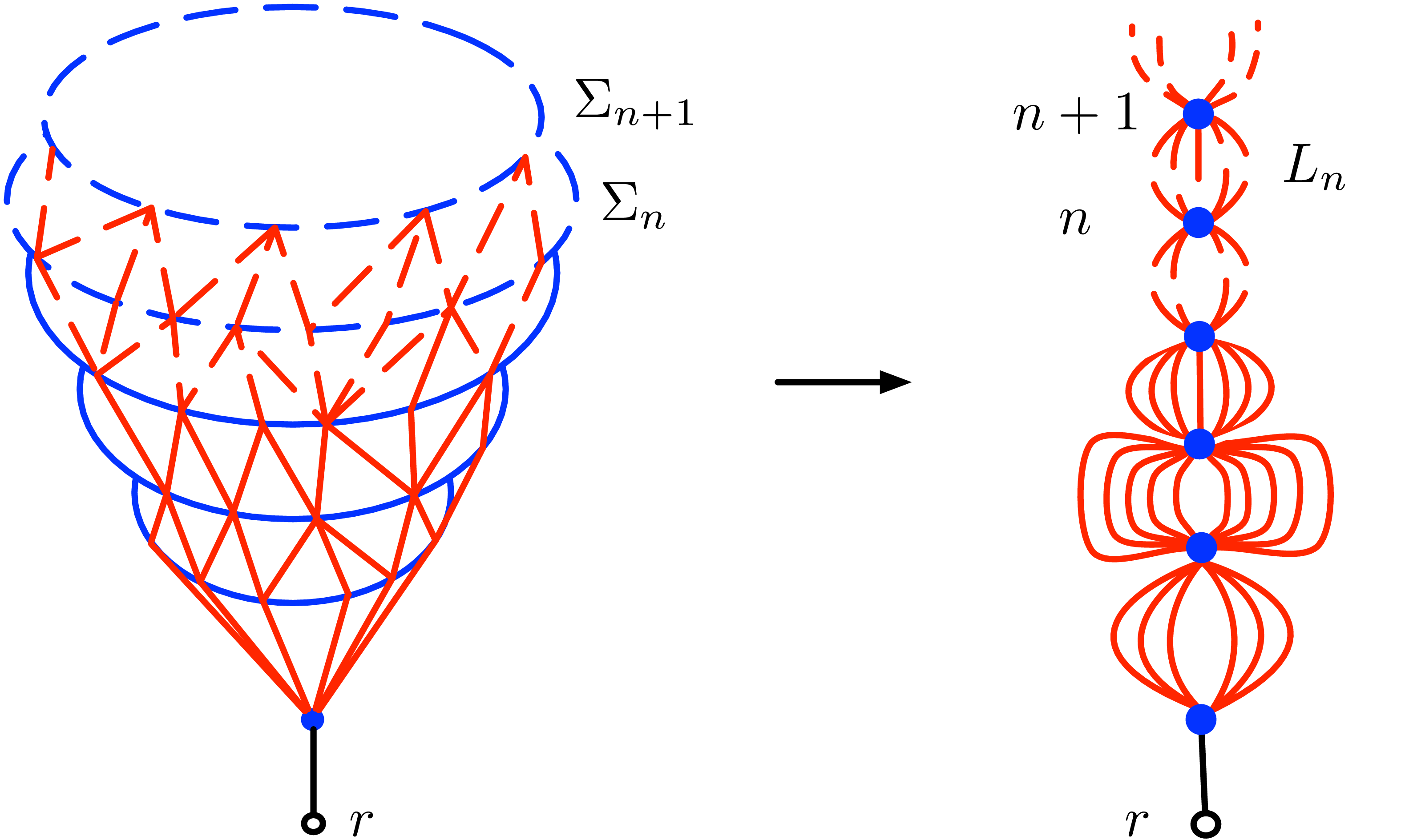}
  \caption{Illustration of how to obtain a multigraph from a causal triangulation.
  }
  \label{fig1}
\end{figure}

The causal constraint in $(d+1)$-dimensional CDT requires the topology to be $I \times \Sigma ^d$ where $d$ is the dimension of spatial hyper-surfaces, which are connected by $(d+1)$-simplices. The foliation structure separates the edges in two types; space-like (blue) and time-like (red) (see Fig.\ref{fig1} for an example in $2+1$ dimensions). Consider the mapping which acts on rooted causal triangulations by collapsing spatial hyper-surfaces into single points (blue) and retaining all the time-like edges (red) (Fig. \ref{fig1}). The resulting graph is a rooted multigraph $M$ with $L_n\geq 1$, $n\geq0$, number of edges connecting vertices $n$ and $n+1$. Note that the multigraph ensemble inherits its measure from the measure of the causal triangulation ensemble. 

We study the spectral dimension $\dS$ on graphs by considering random walks on them and determining the asymptotic behaviour of the first diverging $k$-th derivative of generating function of the return probabilities to the root, $p_G(t)$ (refer to \cite{Giasemidis:2012rf, Giasemidis:2012qk} for details), 
\beq
Q^{(k)}_G(x) \equiv \left ( \sum _{t=0}^{\infty} p_G(t) (1-x)^{t/2}\right )^{(k)} \sim x^{-(k+1)+\dS/2}, \qquad x \to 0, \qquad k\geq 0,
\eeq
for $2k \leq \dS < 2 (k+1)$. Denoting by $M_n$ the multigraph obtained from $M$ by removing all the vertices and edges from the root to vertex $n-1$ and relabelling the remaining multigraph and setting $\eta_{n}\equiv Q_{M_n} /L_n$, the generating function satisfies the recursion relation \cite{Durhuus:2009sm}
\beq \label{rec_rel}
\eta _{M_n} (x) = \eta _{M_{n+1}} (x) + \frac{1}{L_n} - x L_n \eta _{M_n} (x)  \eta _{M_{n+1}} (x).
\eeq
which is the starting point to determine the spectral dimension.

The reason why multigraphs serve as realistic models that share the same spectral dimension with CDT is twofold; firstly it is proven \cite{Durhuus:2009sm} that recurrent multigraphs bound above the spectral dimension of two-dimensional CDT and further this bound is believed to be tight. Secondly, in a rigorous proof \cite{Durhuus:2009sm} it was shown that the number of space-like links in a slice of a two-dimensional CDT remains finite for finite time. This indicates that a random walker on a CDT spends only a finite time on (finite) spatial hyper-surfaces which implies that the spectral dimension is not affected by random walks on spatial surfaces. This argument suggests that the multigraph approximation does not effect the spectral dimension of CDT. 

Having motivated the use of multigraphs, we now discuss the properties that determine the spectral dimension on these graphs. Rigorous results for multigraph ensembles with recurrent measures related to two-dimensional CDT \cite{Durhuus:2009sm, Giasemidis:2012rf} and transient measures \cite{Giasemidis:2012rf} indicate that the spectral dimension is determined by the volume and resistance growth (considering the graph as an electrical network where each edge has resistance one). The knowledge of these two quantities is sufficient to determine the spectral dimension of multigraph ensembles. In the next section we apply this intuition to ``radially reduced'' higher-dimensional CDT ensembles.

\section{Higher-dimensional models and results}
\label{higher_dim}
Due to the absence of an analytical model for higher-dimensional CDT we propose a measure $\mu$ of the corresponding multigraph ensembles by imposing three assumptions \cite{Giasemidis:2012rf, Giasemidis:2012qk}, which are closely related to volume and resistance growth. The first assumption concerns the connectivity, %
\bea \label{connectivity_ansatz}
\left \langle L_N \right \rangle _{\mu (\nu_D)} \simeq \nu_D N^{3-\epsilon} +N.
\eea
Here $\nu_D$ is the inverse bare Newton's constant and $D=d+1$ the topological dimension. In the following we choose $\epsilon=1$ for $D=3$ while for $D=4$, $\epsilon$ is arbitrarily small and is introduced for purely mathematical reasons \cite{Giasemidis:2012rf}. 
The assumption  \eqref{connectivity_ansatz} implies for the volume growth
$\underline{c}N\avg{L_N}_{\mu(\nu_D)} \leq \avg{B(N)}_{\mu(\nu_D)} \leq  \bar{c}N\avg{L_N}_{\mu(\nu_D)} $
and is in agreement with computer simulations of four-dimensional CDT.

The second assumption bounds from above the resistance from the vertex $N$ to infinity and controls the downward fluctuations in connectivity while the third ansatz controls the upward fluctuations in connectivity. In particular, these two assumptions read
\bea \label{resistance_ansatz}
R(N) &\leq & \frac{N}{ \avg{L_N}_{\mu(\nu_D)}}  \psi _{+}(\sqrt{\nu_D} N^{\frac{2-\epsilon}{2}}), \\
\label{connectivity_upper_bound}
L_N &\leq& \avg{L_N}_{\mu(\nu_D)} \psi (\sqrt{\nu_D} N^{\frac{2-\epsilon}{2}})
\eea 
for $N>N_0 >0 $ and almost all graphs of the ensemble. Here $\psi(x),\psi_+(x)$ are diverging and slowly varying functions at $x=0$ and $x=\infty$. Note that \eqref{connectivity_ansatz}-\eqref{connectivity_upper_bound} are higher-dimensional analogues of the analytical two-dimensional model. The description of the higher-dimensional model ($2 \leq \dS \leq 4$) requires transient multigraphs with finite resistance $R(N)$. Therefore the spectral dimension is extracted from $Q'(x)$, which diverges for $x\to 0$.  

The scale dependent spectral dimension depends on a length scale which is set by  $\nu_D$. Special attention has to be paid to define the correct scaling limit of the discrete random graph model in which both the walk length as well as the characteristic length scale diverge in a ``double scaling limit''  \cite{Giasemidis:2012rf, Atkin:2011ak}, 
\beq
\abs{ \tilde Q'(\xi, G_{(D)}) }\equiv \lim _{a \rar 0 } \left(a/G_{(D)}\right) \avg{\abs{Q_M' (x=a \xi)}}_{\mu\left (\nu_D=a^{1-\epsilon/2}/G_{(D)}\right)}.
\eeq
In particular, one obtained for $D=3$ and $D=4$ respectively,
\beq \label{Qprime_cont_Ddim}
 \abs{ \tilde Q'(\xi, G_{(3)}) } \sim
\begin{cases}
\xi ^{-1}, &\xi >> G^{-2}_{(3)},\\
\xi ^{-1 /2}, &\xi <<  G^{-2}_{(3)},
\end{cases} \qquad 
\abs{ \tilde Q'(\xi, G_{(4)}) } \sim
\begin{cases}
\xi ^{-1}, &\xi >> G^{-1}_{(4)},\\
\xi ^{-\epsilon /2}, &\xi << G^{-1}_{(4)},
\end{cases}
\eeq
where $G_{(3)}$ and $G_{(4)}$ are the renormalised Newton's constants in three and four dimensions respectively. This result implies that for $D=3,4$ one has $d_s^{\infty}=D$ at large distances while $d_s^{0} = 2$ at short scales, which agrees with the numerical results in three \cite{Benedetti:2009ge, Kommu:2011wd} and four-dimensional CDT \cite{Ambjorn:2005db, Kommu:2011wd}. Although $G_{(3)}^2$ and $G_{(4)}$ set a scale on the {\it duration} of the walk, it is their square root which corresponds to the {\it length extent} on the graph and is identified with the Planck length in three and four dimensions respectively. 

We can apply a Tauberian theorem to $\avg{\abs{Q'(x)}}_{\mu(\nu_D)}$ to find the average return probability density. Scaling the latter we determine the probability density $P(\sigma)$ for continuous diffusion time $\sigma$. In \cite{Giasemidis:2012qk} we pursued it for $D=4$  and showed that the functional form of $P(\sigma)$ is consistent with the one conjectured from computer simulations  in \cite{Ambjorn:2005db}. Here we give the analogous result for $D=3$. In this case one has
\beq
P(\sigma)  \equiv \lim_{a\to 0} a^{-1} \avg{p(t=\lfloor \sigma/a) \rfloor)}_{\mu(\nu_3)} \simeq \frac{1}{\sigma^{3/2}\left (1+\sqrt{2} G_{(3)}/\sqrt{\sigma}\right )}
\eeq 
 which implies that the scale dependent spectral dimension is given by
 \beq \label{Ds_2+1}
 D_s(\sigma) \equiv -2\frac{d \log P(\sigma)}{d \log \sigma} = 3 - \frac{1}{1+ \sqrt{\sigma/(2 G_{(3)}^2)}}
 = \begin{cases}
 2, \qquad &\sigma \to 0, \\
 3, \qquad &\sigma \to \infty.
 \end{cases}
 \eeq
We note that, while having the correct limits, this result is slightly different than the rational or exponential fit suggested in \cite{Benedetti:2009ge, Kommu:2011wd} to explain the numerical data. It would be interested so to test how well the data fits in the functional form \eqref{Ds_2+1}
 of the spectral dimension.

 \section{Discussion}
\label{discussion}
We reviewed the results of \cite{Giasemidis:2012rf, Giasemidis:2012qk} regarding a radial reduction of CDT, the multigraph approximation, which can be used to study the phenomenon of scale dependent spectral dimension analytically. Besides discussing previous results on the four-dimensional case, we also applied this approximation to three-dimensional CDT.

The three-dimensional model provides an interesting playground to study the relations between different approaches to Quantum Gravity \cite{Anderson:2011bj, Sotiriou:2011aa}. One possibility to relate the results from the multigraph approximation with other approaches is to follow the techniques of \cite{Sotiriou:2011aa} to find the dispersion relation $\omega (k)$ which corresponds to \eqref{Ds_2+1}.
In \cite{Sotiriou:2011mu} a dispersion relation was introduced which originates from the foliation-defining scalar of 2+1 dimensional Ho\v rava-Lifshitz gravity and which is given by 
$\omega(k) ^2 = A k^2\frac{1+B k^2+C k^4}{1+D k^2}$.	
It is readily seen that it has the same asymptotic behaviour as the dispersion relation $\omega(k)$ derived from \eqref{Ds_2+1} when $k \to 0$ and $k \to \infty$. 
Trying to link those two results might shed further light on the relationship between Ho\v rava-Lifshitz gravity and CDT(-like) models.

\ack{This work was supported by the A.G. Leventis Foundation, the A.S. Onassis Foundation grant F-ZG/097/2010-2011, EPSRC grant EP/I01263X/1 and STFC grant ST/G000492/1.}

\section*{References}
\bibliographystyle{iopart-num}

\providecommand{\newblock}{}
\begin{thebibliography}{1}
\expandafter\ifx\csname url\endcsname\relax
  \def\url#1{{\tt #1}}\fi
\expandafter\ifx\csname urlprefix\endcsname\relax\def\urlprefix{URL }\fi
\providecommand{\eprint}[2][]{\url{#2}}

\bibitem{Giasemidis:2012rf}
Giasemidis G, Wheater J~F and Zohren S 2012  (\textit{Preprint}
  \eprint{1202.6322})

\bibitem{Calcagni:2011sz}
Calcagni G 2012 {\em JHEP\/} {\bf 1201} 065 (\textit{Preprint}
  \eprint{1107.5041})

\end{thebibliography}


\begin{thebibliography}{30}
\expandafter\ifx\csname url\endcsname\relax
  \def\url#1{{\tt #1}}\fi
\expandafter\ifx\csname urlprefix\endcsname\relax\def\urlprefix{URL }\fi
\providecommand{\eprint}[2][]{\url{#2}}

\bibitem{Ambjorn:1998xu}
Ambj{\o}rn J and Loll R 1998 {\em Nucl.Phys.\/} {\bf B536} 407--434
  (\textit{Preprint} \eprint{hep-th/9805108})


\bibitem{Ambjorn:2005db}
Ambj{\o}rn J, Jurkiewicz J and Loll R 2005 {\em Phys.Rev.Lett.\/} {\bf 95}
  171301 (\textit{Preprint} \eprint{hep-th/0505113})

\bibitem{weinberg} S. Weinberg, 
in General Relativity, S. Hawking and W. Israel, eds., p. 790-831, 1980.

\bibitem{Litim:2003vp}
Litim D~F 2004 {\em Phys.Rev.Lett.\/} {\bf 92} 201301 (\textit{Preprint}
  \eprint{hep-th/0312114})

\bibitem{Lauscher:2005qz}
Lauscher O and Reuter M 2005 {\em JHEP\/} {\bf 0510} 050 (\textit{Preprint}
  \eprint{hep-th/0508202})

\bibitem{Horava:2009uw}
Ho{\v r}ava P 2009 {\em Phys.Rev.\/} {\bf D79} 084008 (\textit{Preprint}
  \eprint{arXiv:0901.3775 [hep-th]})

\bibitem{Horava:2009if}
Ho{\v r}ava P 2009 {\em Phys.Rev.Lett.\/} {\bf 102} 161301 (\textit{Preprint}
  \eprint{arXiv:0902.3657 [hep-th]})

\bibitem{Modesto:2008jz}
Modesto L 2009 {\em Class.Quant.Grav.\/} {\bf 26} 242002 (\textit{Preprint}
  \eprint{arXiv:0812.2214 [gr-qc]})

\bibitem{Caravelli:2009gk}
Caravelli F and Modesto L 2009  (\textit{Preprint} \eprint{arXiv:0905.2170
  [gr-qc]})

\bibitem{Magliaro:2009if}
Magliaro E, Perini C and Modesto L 2009  (\textit{Preprint}
  \eprint{arXiv:0911.0437 [gr-qc]})

\bibitem{Modesto:2009qc}
Modesto L and Nicolini P 2010 {\em Phys.Rev.\/} {\bf D81} 104040
  (\textit{Preprint} \eprint{arXiv:0912.0220 [hep-th]})

\bibitem{Calcagni:2010pa}
Calcagni G 2011 {\em Phys.Lett.\/} {\bf B697} 251--253 (\textit{Preprint}
  \eprint{arXiv:1012.1244 [hep-th]})

\bibitem{Calcagni:2012rm}
Calcagni G 2012  (\textit{Preprint} \eprint{arXiv:1205.5046 [hep-th]})

\bibitem{Reuter:2011ah}
Reuter M and Saueressig F 2011 {\em JHEP\/} {\bf 1112} 012 (\textit{Preprint}
  \eprint{arXiv:1110.5224 [hep-th]})

\bibitem{Ambjorn:2010hu}
Ambj{\o}rn J, Gorlich A, Jordan S, Jurkiewicz J and Loll R 2010 {\em Phys.Lett.\/}
  {\bf B690} 413--419 (\textit{Preprint} \eprint{arXiv:1002.3298 [hep-th]})

\bibitem{Anderson:2011bj}
Anderson C {\em et~al.\/}
  2012 {\em Phys.Rev.\/} {\bf D85} 044027 
  (\textit{Preprint} \eprint{arXiv:1111.6634 [hep-th]})

\bibitem{Giasemidis:2012rf}
Giasemidis G, Wheater J~F and Zohren S 2012 
{\em J. Phys. A: Math. Gen.} {\bf 45} 355001
 (\textit{Preprint}
  \eprint{arXiv:1202.6322 [hep-th]})

\bibitem{Giasemidis:2012qk}
Giasemidis G, Wheater J~F and Zohren S 2012
{\em Phys.Rev.\/} {\bf D} in press  (\textit{Preprint}
  \eprint{arXiv:1202.2710 [hep-th]})

\bibitem{Durhuus:2009sm}
Durhuus B, Jonsson T and Wheater J~F 2006 {\em J. Stat. Phys.\/} {\bf 139}
  859--881 (\textit{Preprint} \eprint{arXiv:0908.3643 [math-ph]})

\bibitem{Atkin:2011ak}
Atkin M~R, Giasemidis G and Wheater J~F 2011 {\em J.Phys.A\/} {\bf A44} 265001
  (\textit{Preprint} \eprint{arXiv:1101.4174 [hep-th]})

\bibitem{Benedetti:2009ge}
Benedetti D and Henson J 2009 {\em Phys.Rev.\/} {\bf D80} 124036
  (\textit{Preprint} \eprint{arXiv:0911.0401 [hep-th]})

\bibitem{Kommu:2011wd}
Kommu R 2012 {\em Class.Quant.Grav.\/} {\bf 29} 105003 (\textit{Preprint}
  \eprint{arXiv:1110.6875 [gr-qc]})

\bibitem{Sotiriou:2011aa}
Sotiriou T~P, Visser M and Weinfurtner S 2011 {\em Phys.Rev.\/} {\bf D84}
  104018 (\textit{Preprint} \eprint{arXiv:1105.6098 [hep-th]})

\bibitem{Sotiriou:2011mu}
Sotiriou T~P, Visser M and Weinfurtner S 2011 {\em Phys.Rev.Lett.\/} {\bf 107}
  131303 (\textit{Preprint} \eprint{arXiv:1105.5646 [gr-qc]})

\end{thebibliography}
\providecommand{\newblock}{}


\end{document}